\documentclass[12pt,a4paper]{article}
\pdfoutput=1
\usepackage[utf8]{inputenc}

\usepackage[a4paper,left=1.3in,right=1.3in,top=1in,bottom=1in,footskip=0.3in]{geometry}
\usepackage{hyperref}
\usepackage{fancyref}
\usepackage{graphics}
\usepackage{enumerate}
\usepackage{epsfig}
\usepackage{amsfonts,commath}
\usepackage{amssymb}
\usepackage{amsmath}
\usepackage{dsfont}
\usepackage{pifont}
\usepackage{bbm}
\usepackage{multirow}
\usepackage{latexsym}
\usepackage{verbatim}
\usepackage{mcite}
\usepackage{cite}
\usepackage{units}
\usepackage[all]{xy}
\usepackage{cancel}
\usepackage{slashed}
\usepackage{tikz}
\usepackage{tocloft}
\usepackage{uniinput}

\newcommand{\be}{\begin{equation}}
\newcommand{\ee}{\end{equation}}
\newcommand{\ben}{\begin{displaymath}}
\newcommand{\een}{\end{displaymath}}
\newcommand{\bea}{\begin{eqnarray}}
\newcommand{\eea}{\end{eqnarray}}

\newcommand{\bean}{\begin{eqnarray*}}
\newcommand{\eean}{\end{eqnarray*}}
\newcommand{\D}{\mathrm{d}}
\newcommand{\db}[1]{\mathrm{D#1}}

\DeclareMathAlphabet{\mathpzc}{OT1}{pzc}{m}{it}

\begin{document}
\pagestyle{plain}
\makeatletter \@addtoreset{equation}{section} \makeatother
\renewcommand{\thesection}{\arabic{section}}
\renewcommand{\theequation}{\thesection.\arabic{equation}}
\renewcommand{\thefootnote}{\arabic{footnote}}
\setcounter{page}{1} \setcounter{footnote}{0}
\begin{titlepage}

\begin{flushright}
UUITP-15/17\\
\end{flushright}

\bigskip
\begin{center}
\vskip 0cm
{\LARGE \bf Black holes as bubbles of AdS} \\[6mm]
\vskip 0.5cm
{\bf U.~H.~Danielsson, G.~Dibitetto  \,and\, S.~Giri}\let\thefootnote\relax\footnote{{\tt ulf.danielsson@physics.uu.se, giuseppe.dibitetto@physics.uu.se, suvendu.giri@physics.uu.se}}\\
\vskip 25pt
{Institutionen f\"or fysik och astronomi, University of Uppsala, \\ Box 803, SE-751 08 Uppsala, Sweden \\[2mm]}
\vskip 0.8cm
\end{center}
\vskip 1cm

\begin{center}
{\bf ABSTRACT}\\[3ex]
\begin{minipage}{13cm}
\small
In this paper we propose that bubbles of AdS within Minkowski spacetime, stabilized at a finite radius by stiff matter and an electromagnetic gas, can be an alternative endpoint of gravitational collapse. 
The bubbles are horizonless with a size up to 12.5\% larger than their Schwarzschild radius depending on their charge.  We argue that they are stable against small perturbations, and have thermodynamical properties similar to those of real black holes. 
We provide a realization of the bubbles within string theory that relies on a specific brane intersection giving rise to a shell carrying dissolved charges from lower dimensional 
D-branes as well as a gas of open strings. We also note that our construction provides a new way of understanding the entropy of Reissner-Nordstr{\"o}m black holes in the extremal limit.
\end{minipage}
\end{center}

\vfill
\end{titlepage}

\tableofcontents
\section{Introduction}\label{sec:introduction}

	Black holes as classical solutions of Einstein gravity pose many puzzles that reveal a profound conflict between quantum mechanics and general relativity \cite{Hawking:1976ra}.    
	By means of semi-classical arguments, one is easily convinced that a  black  hole  possesses an  entropy, which is given by its horizon area in Planck units \cite{Hawking:1976de},
	while classically, in general relativity, a black hole solution turns out to be unique for a given value of its mass, charge and angular momentum \cite{Ruffini:1971bza}.  
	Such a \emph{no-hair} theorem appears, then, to be in contrast with the existence of any microscopic description of a given black hole, at least at a classical level.
	Due to the above puzzle, the issue of constructing microstates of a black hole properly accounting for its entropy is naturally turned into one of the biggest challenges for a theory of quantum 
	gravity. 
	
	Moreover, the enormous black hole entropy which is not visible at the black hole horizon causes a violation of unitarity, in such a way that the information concerning the original black hole state 
	cannot be encoded into the  Hawking  radiation. Therefore, any resolution of this puzzle requires new physics at the horizon scale \cite{Mathur:2009hf}. 
	However, due to the horizon being a null surface, it turns out to be impossible to classically add new structure at that scale, in that any form of matter will either fall into the singularity or dilute
	very quickly.
	
	In this context, black hole complementarity \cite{tHooft:1991uqr,Susskind:1993if} was proposed as a way of reconciling the non-unitary phenomenon of black hole evaporation through Hawking radiation with string theory, 
	as a proposal for a unitary theory of quantum gravity. This idea suggests that information is both reflected at the event horizon and transmitted through without being able to escape, in such a way that
	no observer can access both simultaneously. As a consequence, nothing special happens at the horizon and all information passes through according to an in-falling observer, while it gets completely absorbed into
	a stretched horizon according to an external observer. 
	
	Nevertheless, whether a black hole really has a horizon, and whether there actually is an interior to fall into, has been increasingly questioned during the last several years.
	The work on firewalls \cite{Almheiri:2012rt} suggests that the idea of black hole complementarity might not work or is at least incomplete. The main inconsistency there being the fact that any outgoing 
	particle would have to be entangled with both its past Hawking radiation and its twin in-falling particle. The firewall resolution of this paradox mainly relies on an immediate breakdown of entanglement
	as soon as the in-falling and outgoing particles get separated on the two opposite sides of the horizon. 
	
	Parallely, the work on fuzzballs \cite{Lunin:2002qf,Giusto:2004id,Mathur:2005zp} suggests that string theory should give rise to a new state of matter that prevents a black hole from forming in the first place.
	According to such a proposal, the underlying black hole microstates consist of wrapped branes yielding a perfectly smooth and horizonless geometry.
	In this case there are different views on what an in-falling observer would actually experience. Some argue that the in-falling observer, even though dissolved into fuzz, should effectively measure 
	something close to what general relativity predicts. Others hold the option open that the journey might end dramatically when the new state of matter is reached.  
	
	In this paper we take this latter possibility seriously in the context of astrophysical black holes. We argue that string theory might replace a Schwarzschild black hole with a bubble of AdS space enveloped 
	by a brane. The matter degrees of freedom live on the brane, and we will be able to show that the thermodynamical properties of the black hole are successfully reproduced by such a \emph{black shell}. 
	Our approach is inspired by the work on gravastars in \cite{Visser:2003ge}. 
	Interestingly, we find a universal prediction for the radius of the shell that is significantly larger than the Schwarzschild radius. We suggest that our construction could be relevant in studies of, 
	\emph{e.g.}, gravitational radiation from colliding black holes.
	
	The paper is organized as follows. In section~\ref{sec:gravastars}, we review the model of a gravastar adapted to the case of an AdS interior and use it to describe (non-) extremal Reissner-Nordstr\"om
	black hole geometries and further discuss stability issues. In section~\ref{sec:bubble}, we discuss the actual probability of nucleating such an AdS bubble within Minkowski spacetime and subsequently keeping 
	it dynamically stable at a fixed radius. In section~\ref{sec:string}, we present a concrete stringy realization of the above ideas by employing a particular brane system in massive type IIA string theory
	consisting of polarized branes wrapping an $S^{2}$ in spacetime and carrying lower-dimensional brane charges in a dissolved form. Finally, we present our conclusions and discuss further possible developments
	in section~\ref{sec:conclusions}.

\section{AdS gravastars}\label{sec:gravastars}
	
	There have been previous attempts to replace actual black holes by other compact objects. General relativity typically requires extreme equations of state in order to stabilize an ultra compact object 
	when attempting to push its size down towards the Schwarzschild radius. 
	Depending on the type of matter that one considers, there is a limit beyond which collapse is inevitable.  For instance, for a spherically symmetric object made of ordinary matter with a 
	density that increases monotonically towards the center, the radius cannot be smaller than $9/8$ times the Schwarzschild radius. This is often called the \textit{Buchdahl bound} \cite{Buchdahl:1959zz}. 
	However, by allowing for exotic matter, the equilibrium radius may be pushed beyond this limit towards an object of smaller radius.
	An example of this is provided in \cite{Visser:2003ge} where the authors assume a thin shell of matter with some mass density and pressure, surrounding a volume of de Sitter space, and find it possible 
	to squeeze the shell arbitrarily close to the Schwarzschild radius. 
	
	In this section we will investigate an especially intriguing possibility that has the benefit of making sense from the point of view of string theory.  Rather than a bubble of de Sitter space, we will consider a 
	bubble of AdS space.\footnote{AdS space was briefly considered in \cite{Visser:2003ge} for a special example.}, the wall separating the AdS interior from outer flat space being composed of branes 
	available in string theory.
	
	\subsection{Black holes as black shells}\label{sec:shell}
		
		Let us first start by trying to get an AdS bubble stabilized at a finite radius, but carrying no electromagnetic charge. This will result in an outer geometry which looks like a neutral Schwarzschild
		black hole geometry.
	
	\subsubsection{Neutral Schwarzschild black hole}\label{sec:schwarzschild}
	
		We consider a shell of matter (of radius $r$) with matter density $\rho$ and two dimensional pressure $p$. Inside the shell we have a cosmological constant $\Lambda <0$, and outside of the shell a Schwarzschild geometry with mass $M$. 
		For stability we require the Israel-Lanczos-Sen \cite{Israel:1966rt,Lanczos:1924,Sen:1924} thin shell junction conditions
		\begin{eqnarray}\label{eq:thinshell}
		\rho &=& \frac{1}{4\pi r} \left(\sqrt{1+k r^2}-\sqrt{1-\frac{2M}{r}}\right)\ ,\\
		p &=& \frac{1}{8 \pi r} \left(\frac{1-\frac{M}{r}}{\sqrt{1-\frac{2M}{r}}} -  \frac{1+2k r^2}{\sqrt{1+k r^2}} \right)\ .
		\end{eqnarray}
		We work in units where $G_{N}=1$ and we have defined $k\,\equiv\,\abs{\frac{\Lambda}{3}}$ with $\Lambda<0$. 
		Using Friedmann's equation in $2+1$ dimensions, pressure can be related to the energy density through the continuity equation. Considering the radius $r$ as a function of time, this is given by
		\begin{equation}\label{eq:continuity}
		\dot{\rho} + \frac{2\dot{r}}{r} \left( \rho +p\right) \,=\,0\ ,
		\end{equation}
		which can be written as
		\begin{equation}
		\partial_r \rho +\frac{2}{r} \left( \rho +p\right) \,=\,0\ ,
		\end{equation}
		or
		\begin{equation}
		p\,=\,-\rho -\frac{r}{2} \frac{d\rho}{dr}\ .
		\end{equation}
		The first of the junction conditions can be viewed as imposing conservation of the total energy when written as 
		\begin{equation}
		4\pi r^2\rho - r \left(\sqrt{1+k r^2}-1\right) =E\ ,
		\end{equation}
		where the two terms appearing on the left hand side represent the energy of the shell and the (negative) energy of the AdS bubble, respectively, while on the right hand side we have the energy of the Schwarzschild black hole,
		which is given by
		\begin{equation} \label{eq:E}
		E\,=\,r-r\sqrt{1-\frac{2M}{r}}\ .
		\end{equation}
		The energies are given relative to the outer empty Minkowski space. The black hole energy includes a gravitational self-interaction term and solves
		\begin{equation}
		M\,=\,E-\frac{E^2}{2r}\ .
		\end{equation}
		The tension of the branes will be set by high energy physics and as a consequence, the size of the negative cosmological constant as well. 
		Expanding for a large cosmological constant, \emph{i.e.} large $k$, and keeping the leading terms we get
		\begin{eqnarray}
		\rho & = & \frac{k^{1/2}}{4\pi}+\frac{1}{8\pi k^{1/2} r^2}  -\frac{1}{4\pi r} +\rho _b \ ,\\
		p & = & -\frac{k^{1/2}}{4\pi}  +\frac{1}{8\pi r} +p _b \ ,
		\end{eqnarray}
		where $\rho _b$ and $p_b$ are defined by comparing with \fref{eq:thinshell} and \fref{eq:E} as
		\begin{eqnarray}
		\rho_b &=& \frac{1}{4 \pi r}\left(1 - \sqrt{1-\frac{2M}{r}}\right)\ ,\\
		p_b &=& \frac{1}{8 \pi r}\left(\frac{1-\frac{M}{r}}{\sqrt{1-\frac{2M}{r}}}-1\right)\ .
		\end{eqnarray}
		Later in the paper we will provide a detailed stringy construction realizing this effective 4D model. Nevertheless, let us briefly go through a heuristic argument that roughly explains how all of this 
		could be understood from string theory. A good starting point is to consider black holes built up from 4 dimensional D-particles. An extremal black hole would consist of the same number of such particles as its charge in fundamental units. A non-extremal one would have pairs of particles and antiparticles in analogy with \cite{Klebanov:1996un,Danielsson:2001xe}. 
		We propose that this is not the whole story but that these D-particles polarize \cite{Myers:1999ps} and become dissolved in the aforementioned spherical branes. 
		The action for the polarized system (reduced to $2+1$-dimensions) is given by
		\begin{equation}
		S\,=\, \int d^3 \sigma \, \tau \, e^{-T^2} \sqrt{ -\,{\rm det} (h_{\mu \nu} +\cal{F}_{\mu \nu} )}\ .
		\end{equation} 
		Schematically, the DBI-action gives an energy $4\pi T_2\sqrt{r^4+n^2}$, where $T_2$ is the effective tension of the 2-brane and $n$ is the number of dissolved D-particles. At $r=0$, we recover $4\pi  T_2 \,n$ as the mass of the D-particles. Conversely, at large $r$ the 2-branes dominate and we get $\left(4 \pi \,T_2 r^2 + \frac{4\pi T_2 n^2}{2r^2}\right)$, with the additional mass due to the D-particles suppressed as $r$ increases. Their contribution to the energy density on the brane goes like $1/r^4$, if $n$ is kept constant, which is the characteristic scaling behavior of two dimensional stiff matter with $p=\rho$.
		
		Note that, in presence of D-particles as well as anti D-particles, two separate terms are needed in the action to account for both. The contribution to ${\cal F}_2$ in the Wess-Zumino-Witten term cancels so that only 
		the net D0-charge appears, while they will add in the tension. Note also the presence of the tachyon field $T$ that allows the brane to vary its effective tension $T_2=\tau e^{-T^2}$ 
		(above a minimum set by the charge it carries).
		
		The important point for us is that this contribution to the tension from the dissolved branes will be almost invisible in our limit of high-energy branes with macroscopic radii. Nevertheless, the presence of the dissolved branes, and the ${\cal F}_2$ field strength on the branes, will play an essential role. In particular, in case of a black hole with a non-vanishing net charge, it is responsible for the coupling of the brane to a spacetime electric field through the Wess-Zumino-Witten term. In addition, it provides the coupling of $n^2$ different kinds of massless open strings to the brane, thereby allowing the existence of a gas at a finite temperature.  
		
\begin{figure}
		\centering
		\includegraphics[width=0.6\textwidth]{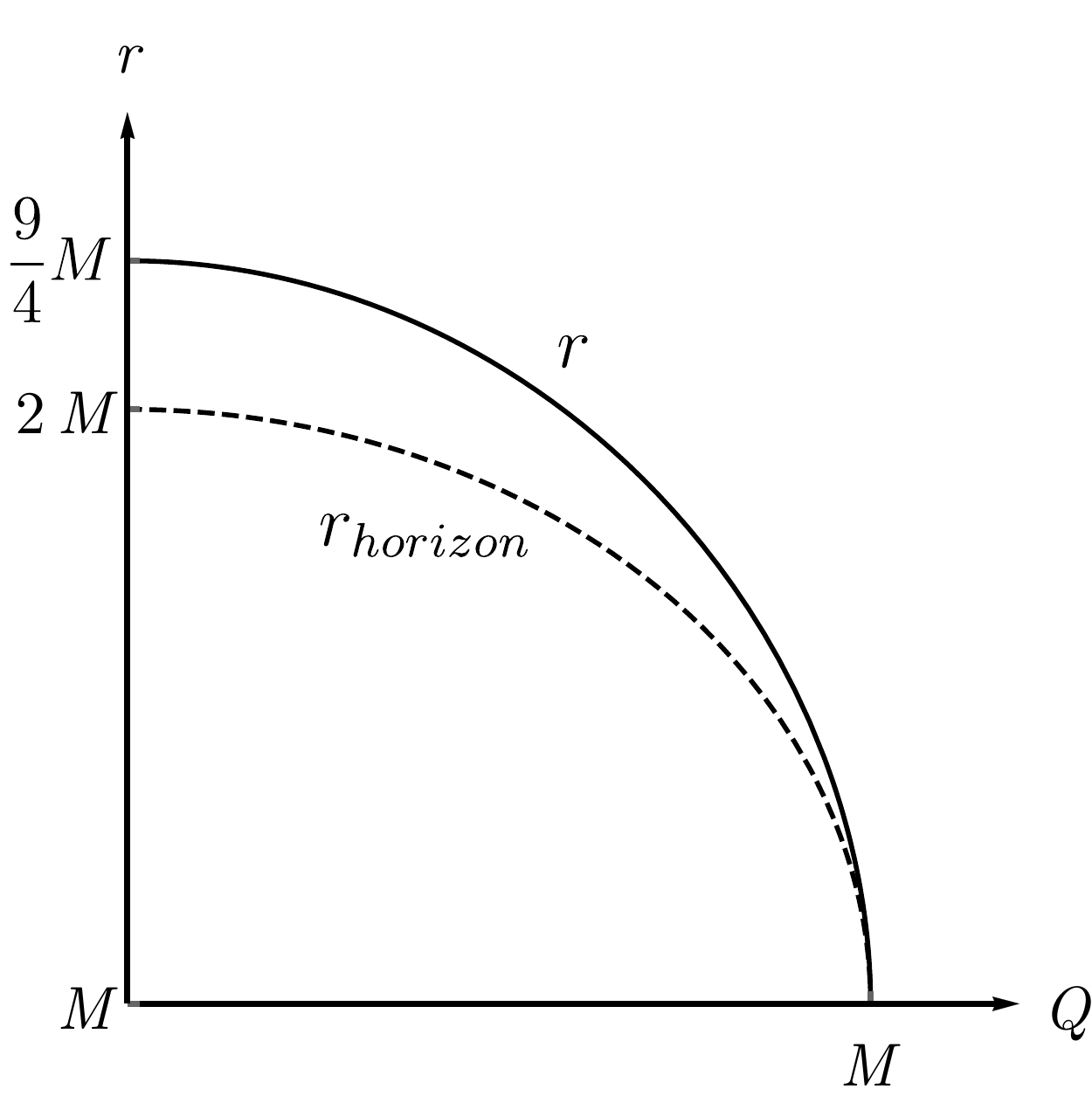}
		\caption{{\it The equilibrium radius of the spherical shell $r$ as a function of the total charge $Q$. This approaches the position of the horizon of an extremal Reissner-Nordström black hole $r_{\rm horizon}$ as $Q\rightarrow M$.}}
		\label{fig:radius}
		\end{figure}

		Summarizing, we claim that the junction conditions take the form
		\begin{eqnarray}
		\tau + \rho _g+ \rho_s & = &\frac{k^{1/2}}{4\pi}+\frac{1}{8\pi k^{1/2} r^2}-\frac{1}{4\pi r} + \rho_b\ ,\\
		-\tau + p_g  + p_s & = & -\frac{k^{1/2}}{4\pi}+\frac{1}{8\pi r}+p_b\ ,
		\end{eqnarray}
		where $p_g= \frac{1}{2} \rho _g$, and $p_s= \rho_s$ . If we assume that neither $\tau$ nor $\rho _s$ depend explicitly on $\rho_b$, the solution is \emph{uniquely} determined and it is given by
		\begin{eqnarray}\label{eq:junction-solution}
		\rho _g & = & \rho_b -\frac{1}{12\pi r}\ ,\\
		\tau & = & \frac{k^{1/2}}{4\pi} - \frac{1}{6\pi r}+\frac{1}{16\pi k^{1/2} r^2}\ , \\
		\rho_s  &= & \frac{1}{16\pi k^{1/2} r^2}\ ,
		\end{eqnarray}
		with $p_b = \frac{1}{2} \rho _b$. \footnote{Note that the formulae determine the required values of $\tau$, $\rho_g$, and $\rho_s$ at a critical point, but not their dependence on $r$ in general.} Remarkably, this uniquely fixes the radius of the shell to the Buchdahl radius at $r=\frac{9M}{4}$. The same above expressions will also hold for non-zero charge, with the charge only appearing explicitly in the expression for $\rho_b$, and hence $\rho_g$. As will be discussed later, the radius of the system will shift, as the charge is increased, from  $r=\frac{9M}{4}$ down to the horizon at $M$ for the extremal case. This situation is depicted in \fref{fig:radius}.
		
		\begin{figure}
		\centering
		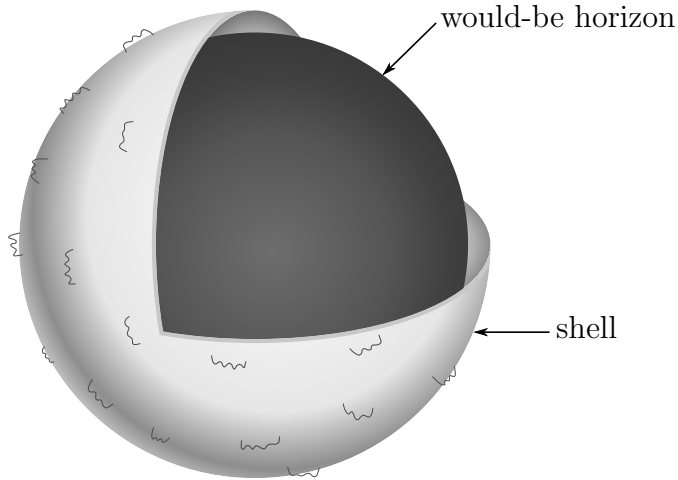
		\caption{{\it An artist's impression showing a cut-away of the microscopic description of the blackshell (light sphere) with a gas of excited open strings on top of it. The dark sphere seen through the cut-away marks the position of the would-be horizon.}}
		\label{fig:shell}
		\end{figure}
		Going back to the stringy picture of our black shell given in terms of polarized branes, let us now be a little bit more specific and consider a D-brane polarized along an $S^2$ in space-time\footnote{This can be compared to the setup in \cite{Kachru:2002gs} where anti-D3 branes polarize into NS5 branes in internal space, whereas we consider a D-brane polarizing in space-time.} with $d$ wrapped internal dimensions of equal size, and let us write $\sqrt{k}=1/R$, where $R$ is the AdS-radius. We then get
		\begin{equation}
		R \,\sim \,\frac{L^{6-d}}{\ell_s^{6-d}} \frac{\ell_s}{g_s}\ ,
		\end{equation}
		where $L$ is the size of the extra dimensions. The density of the stiff matter is fixed so that
		\begin{equation}
		\frac{n^2 \ell_s^4}{r^4} \,\sim \,\frac{R^2}{r^2} \ ,
		\end{equation}
		implying
		\begin{equation}
		n \,\sim \, \frac{rR}{\ell_s^2} \,\sim \,\frac{L^{3-d}}{\ell_s^{3-d}} \frac{r}{\ell_{\rm Pl}}\ .
		\end{equation}
		With exactly three wrapped dimensions, \emph{i.e.} a D5-brane, we find $n^2 \,\sim \,\frac{r^2}{\ell_{\rm Pl}^2}$, which is the expected number of degrees of freedom. This guarantees that the energy $ 4 \pi r^2 \rho _g \,\sim \,M$ with $\rho _g \,\sim\, n^2 T^3$, if $T \sim 1/r$. This result is invariant under S- and T-duality since it only depends on $\ell_{\rm Pl}$. For instance, in case of a D3 polarizing into an NS5, the DBI-action has an overall $1/{g_s^2}$, and an extra ${g_s^2}$ in front of the $n^2$, which leads to the same result. In section 4 we will see how this works out in detail for a triple T-dual type IIA configuration.
	
	\subsubsection{Stability against small perturbations}
		
		At first we start out by testing what happens if there is no energy transfer  between the different components in the system. Following the analysis carried out in \cite{Visser:2003ge}, we write
		\begin{equation}
		\rho \,=\, \frac{1}{4\pi r} \left(\sqrt{1-2V(r)+k r^2}-\sqrt{1-2V(r)-\frac{2M}{r}}\right)\ .
		\end{equation}
		The junction conditions then become equivalent to stationary solutions of the system
		\begin{equation}
		\frac{\dot{r}^2}{2} + V(r) \,=\,E\ ,
		\end{equation}
		\emph{i.e.} profiles of the form
		\begin{equation}
		\begin{array}{lclcclc}
		\dot{r}=0 & \textrm{at constant} & r=r_0 & , & \textrm{and} & E \ = \ V(r_0) \ = \ V'(r_0) \ = \ 0 & .
		\end{array}
		\end{equation}
		Stability can be checked by taking yet another derivative with respect to $r$ and adjusting the second derivative of $V(r)$ such that we get the expected behavior for $\partial_{rr}\rho$ assuming that $\rho$ has contributions coming from both a brane with constant tension $\tau$ and a gas with $\rho _g \sim 1/r^3$. Working through these steps one easily concludes that $\partial_{rr}V<0$. As a consequence, the shell is unstable with respect to small perturbations, and will start to move away from the critical point either by contracting or by expanding. Equivalently, one can simply observe that the pressure of the shell, when its radius is reduced, is {\it smaller} than what is required by the junction condition for stability, and the shell will therefore be pushed to an even smaller radius. Vice versa, if the radius is increased beyond the critical point. Hence we seem to conclude that the configuration is unstable.
		
		However, a more careful analysis is required in order to assess whether this is really what happens. We have a gas consisting of $n \sim r$ massless particles at a temperature $T \sim 1/r$, yielding an energy density $\rho \sim n^2 T^3 \sim 1/r$. When $n$ is assumed to be constant we get $\rho _g \sim 1/r^3$. But what is the temperature and what is its origin?  Naively, one might be tempted to invoke the local Hawking temperature $T_{\rm H}=\frac{1}{8\pi M \sqrt{1-\frac{2M}{r}}}$, aiming to reproduce the exact same thermal properties as the ones of a corresponding black hole carrying the same mass. However, there is no real reason for doing so. Instead, the natural temperature is the local Unruh temperature \cite{Unruh:1982ic}
		\begin{equation}
		T_{\rm U}=\frac{a}{2\pi} = \frac{M}{2\pi r^2 \sqrt{1-\frac{2M}{r}}}\ ,
		\end{equation}
		where $a$ is the proper acceleration of the shell.
		The correct vacuum is picked by studying the process that forms the object. In the case of a black hole that is the result of a collapsing star, the infalling Minkowski vacuum develops into the Hawking vacuum at finite temperature, rather than to the zero temperature Boulware vacuum. Similarly, when our shell forms it will find itself accelerating with respect to the infalling Minkowsky vacuum, suggesting that it will be heated to the Unruh temperature. Our conclusion agrees with \cite{Saravani:2012is}, where the same choice of vacuum was made in a similar situation. It is reasonable to assume that the gas on top of the shell is heated to the same temperature.
		It is only when $r \rightarrow 2M$ that $T_{\rm U} \rightarrow T_{\rm H}$, and it is lower otherwise. A well known way to argue for the Hawking temperature, is indeed to support the microscopic degrees of freedom of the black hole on a membrane which is placed at a Planck length or so away from the horizon. The local Unruh temperature will be red shifted to the Hawking temperature far away. We will argue in just the same way, but assume our shell to be much further away from the would-be horizon, as shown in \fref{fig:shell}. 
		In particular, at $r=\frac{9M}{4}$ we find the local temperature to be given by $T_{\rm U}=\frac{8}{27\pi M}$. If this is the temperature of the black shell in its local rest frame, the temperature measured by an asymptotic observer (at $r \rightarrow \infty$) will be $\frac{8}{81\pi M} < \frac{1}{8\pi M}$.
		
		If we were to compress the shell, at constant $n$, and with no extra transfer of the energy to the gas, $\rho _g$ would simply respond as $\rho _g \sim 1/r^3$. On the other hand, if we assume that the temperature adjusts itself to the new Unruh temperature the situation will be completely different. Some energy then needs to be transferred to the gas, and the only possible source is the brane\footnote{The scenario is exactly the same as was proposed in \cite{Danielsson:2004xw} in the context of non-Bunch-Davies vacua in an inflationary cosmology. There, particle creation depletes the cosmological constant and leads to running.}.
		
		Let us see how this works in detail. We work to lowest order, and neglect the subleading contribution from the stiff gas. To this end, we split the continuity \fref{eq:continuity} into two parts, one for the brane and one for the gas:
		\begin{eqnarray}
		\dot{\tau} &=& -j  \ ,\\
		\dot{\rho _g} + \frac{3\dot{r}}{r} \rho_g  &=&j\ ,
		\end{eqnarray}
		where $j$ is a source term. This can also be written as
		\begin{eqnarray}
		\partial_r \tau &=&- \frac{j}{\dot{r}}\ , \\
		\partial_r \rho_g + \frac{3}{r} \rho_g &=&\frac{j}{\dot{r}}\ .
		\end{eqnarray}
		The source term $j$ represents the energy transfer that adjusts the temperature of the gas so that it follows the Unruh temperature. It cancels out in the expression of the total energy, thus correctly accounting for an energy transfer. Varying the first junction condition, assuming that it always holds, we get
		\begin{equation}
		\partial_r \tau + \partial_r \rho_g\,=\, -\frac{4}{81\pi m^2}\ .
		\end{equation}
		Assuming further that $n$ be unaffected as the brane changes its temperature, we use $\frac{\partial_r T}{T}=-\frac{8}{3M}$ to get $\partial_r \rho_g = 3\frac{\partial_r T}{T} \rho_g = -\frac{8}{27\pi m^2}$. This allows us to determine the change in the brane tension to be
		\begin{equation}
		\partial_r \tau \, =\,\frac{20}{81\pi m^2}\ .
		\end{equation}
		That is, the tension of the brane reduces when $r$ is decreased. We now find:
		\begin{equation}
		\partial _r p \,=\, -\partial_r \tau +\frac{1}{2} \partial _r \rho _g \,=\, -\frac{32}{81\pi m^2} \,<\, -\frac{14}{81\pi m^2}\ ,
		\end{equation}
		where we have compared with the derivative of the second junction condition. We conclude that if the shell is compressed, the pressure of the shell becomes {\it larger} than what the junction condition requires for stability, and the shell is pushed back out. Vice versa for a shell at a larger radius. Physically, this is just what one would expect. When the shell is compressed, the gas is heated up and wants to be pushed back out. Similarly, energy is depleted from the brane that relaxes its grip and lets the shell move back out. 
		
		Note that, in the above argument, we have assumed that $n$ does not change. If we compress the brane, heating up the system, one would at least naively expect $n$, the number of dissolved brane/anti-brane pairs, to increase. This would increase the energy of the gas even further, in favor of stability. On the other hand, if $n \sim r$ as is the case at equilibrium, the energy increase of the gas will be somewhat reduced. Nevertheless, it is easy to check that the system will still be stable.

		In our analysis we have ignored finite size effects and possible effects due to strong coupling. A full analysis would require a better understanding of the detailed dynamics of the gas, and the other matter components on the shell. With this caveat, we conclude that the shell is \emph{stable} under small perturbations provided that the gas is allowed to adjust itself to the Unruh temperature.
	
	\subsection{(Non-)extremal Reissner-Nordstr{\"o}m  black hole}\label{sec:RN-gravastar}
	
		Let us now move to considering a black shell carrying some net electromagnetic charge, thus effectively describing an outer Reissner-Nordstr{\"o}m  black hole geometry.
		The construction works similarly to the previous case, at least in spirit. The junction conditions are now given by
		\begin{eqnarray}
		\rho &=& \frac{1}{4\pi r} \left(\sqrt{1+k r^2}-\sqrt{1-\frac{2M}{r}+\frac{Q^2}{r^2}}\right) \ , \\
		p &=& \frac{1}{8 \pi r} \left(\frac{1-\frac{M}{r}}{\sqrt{1-\frac{2M}{r}+\frac{Q^2}{r^2}}}-\frac{1+2k r^2}{\sqrt{1+k r^2}} \right)\ ,
		\end{eqnarray}
		and the radius of the shell solves
		\begin{equation}
		1-\frac{M}{r}\,=\,2 f(r)^{1/2}-f(r)\ ,
		\end{equation}
		where $f(r)\,\equiv\,1-\frac{2M}{r}+\frac{Q^2}{r^2}$. The local Hawking temperature is given by 
		\begin{equation}
		T_{\rm H}\,=\,\frac{\kappa (r_+)}{2\pi}\frac{1}{f(r)^{1/2}}\ ,
		\end{equation}
		where
		\begin{equation}
		\kappa (r) \,=\,\frac{1}{2} f'(r)\,=\,  \frac{M}{r^2} -\frac{Q^2}{r^3} \ ,
		\end{equation}
		is the surface gravity at radius $r$.  In particular, $\kappa (r_+) =\frac{r_+ - r_-}{2 r_+^2}$ where $r_{\pm} \equiv M \pm \sqrt{M^2-Q^2}$.
		For us the relevant temperature is again the Unruh temperature, which is in turn given by
		\begin{equation}
		T_{\rm U}\,=\,\frac{\kappa (r)}{2\pi}\frac{1}{f(r)^{1/2}}\ ,
		\end{equation}
		evaluated at the radius of the shell.
		
		Let us now focus on the near-extremal limit for a Reissner-Nordstr{\"o}m black hole. It may be seen that the aforementioned limit is approached by taking
		\begin{eqnarray}
		\frac{M}{r} &=& 1- \epsilon_1 \ ,\\
		\frac{Q^2}{r^2} &=& 1 -\epsilon _2\ ,
		\end{eqnarray}
		where $\epsilon _2 = 2\epsilon _1 - \frac{\epsilon _1^2}{4}$ such that  $p_g = \frac{1}{2} \rho _g$. The surface gravity vanishes in the extremal limit, but the blow up of the blueshift as the horizon is approached still turns out to yield a finite temperature given by
		\begin{equation}
		T_{\rm RN} \,=\, \frac{1}{\pi M}\ .
		\end{equation}
		
		Interestingly, we find that the entropy of the black hole in the extremal limit can be carried by a gas at non-zero temperature. In this way we hope to have clarified a long-standing confusion concerning the possibility for an extremal Reissner-Nordstr\"om black hole to carry non-vanishing entropy.
		The confusion arises from the fact that, while a semiclassical calculation would seem to indicate that such an object should have vanishing entropy, the area of its event horizon is non-zero 
		\cite{Hawking:1994ii,Teitelboim:1994az} (see also \cite{Carroll:2009maa}).

		In this context one should note that there is an alternative way of solving the junction conditions in case of the extremal Reissner-Nordström black hole. Just assume a shell enclosing a region of flat space with zero cosmological constant, with the junction conditions collapsing to
		\begin{eqnarray}
		\rho &=& \frac{Q}{4\pi r^2}\ , \\
		p&=&0\ ,
		\end{eqnarray}
		where $Q=M$. This is simply a shell of pressure-less dust that can be put at any radius outside of the horizon, and is a simple consequence of the cancellation of the gravitational and electric forces between the particles. If we take the limit of a large number of particles we get a continuous shell with a metric without any singularities. The mass of the black hole is fully carried by the D-particles, and there is no need for a gas. We think that this latter construction with dust could be closer to the fuzzballs of \cite{Bena:2016ypk,Bena:2017geu} than the ones with branes above. 
		The goal in these papers was to construct horizonless black holes using a finite number of particles. This can be achieved if the particles in the multi-centered solutions are carefully positioned just at the right places. Our continuum limit is of course not sensitive to these details. This kind of extremal black hole cannot be obtained through a limit of the non-extremal case, which, we suggest, leads to the inevitable presence of branes in which the dust is dissolved.
		
		To our understanding, there exist, therefore, two different microscopic descriptions of an extremal black hole, one of which describes a supersymmetric system, whereas the other one does not. 
		In one description the extremal black hole consists of charged dust at zero temperature, its (possibly) non-vanishing value of the entropy simply accounting for a non-trivial degeneracy index of the vacuum state of the system. This is the result which was first successfully reproduced in \cite{Strominger:1996sh} and subsequently in many other works in various other cases along the same lines.
		The other description instead retains a gas at finite temperature while taking the near-extremal limit of a non-extremal black hole. Far away from the black hole the temperature approaches zero, but just at the horizon a finite value remains in that limit. It is this gas that carries the entropy. 
		For this mechanism to actually work, it is essential that the contribution to the mass due to the elementary charges, or D-particles, be suppressed and effectively vanishing. 
		As explained previously, this comes about since they are dissolved in the high tension brane.

\section{Bubble nucleation}\label{sec:bubble}

	So far we have managed to establish the existence of ultra compact objects in the form of black bubbles of AdS space. There are now two important things we need to do. First, we need to check the stability of the Minkowski vacuum against spontaneous and disastrous formation of bubbles leading to a phase transition. Second, we need to show that stable bubbles are likely to form at the end of gravitational collapse.
	
	The probability of tunneling can be obtained by integrating the junction condition corresponding to energy conservation. Following the analysis initiated in \cite{Brown:1988kg}, and further expanded in \cite{HenryTye:2008xu},
	we write the junction condition between AdS space and Minkowski as 
	\begin{equation}
	\frac{\partial B}{\partial r} \,= \,6 \pi^2 r^2 \left( \rho - \frac{1}{4 \pi r} \left(\sqrt{1+k r^2}-1 \right) \right) \,=\,0\ ,
	\end{equation}
	where $B$ is the instanton action, and the probability of tunneling can be written $\sim e^{-B}$. Integrating, and fixing the constant of integration so that $B$ vanishes at $r=0$, we find
	\begin{equation}
	B=2 \pi^2 \rho r^3 - \frac{\pi}{2} \left( \frac{\left( 1+k r^2 \right) ^{3/2} -1}{k}-\frac{3r^2}{2} \right)\ .
	\end{equation}
	Here we have assumed that  $\rho$ is a constant representing pure tension.
	
	If we evaluate the instanton action at its extremum, \emph{i.e.}, when the junction condition is satisfied, we find
	\begin{equation}
	B\,=\, \frac{\pi r^2}{4} +\frac{\pi}{2k} \left( 1 -  \sqrt{1+kr^2} \right)\ .
	\end{equation}
	The actual value is set by the tension $\rho$ of the brane, and the cosmological constant within the bubble. With the AdS-radius much larger than the Planck scale it follows that  $B$ is always of order $r^2$ in Planck units. If this radius is at least a few orders of magnitude larger than the Planck scale, the formation of bubbles is heavily suppressed. In the limit which is relevant to us, the radius is much larger than the AdS-radius, and therefore $B \sim  \frac{\pi r^2}{4}$ (in Planck units).
	
	A natural question that may arise at this point concerns tunneling during gravitational collapse.  Let us now assume, for simplicity, that the collapsing star is in the form of a thin shell of matter with Schwarzschild geometry on the outside, and Minkowski space on the inside. As we have seen, the formation of an AdS-bubble somewhere inside of the collapsing shell will be heavily suppressed, unless it lands right on top of the shell. In such a case, it can then immediately absorb all of the matter content, and transform it into brane/anti-brane pairs supporting a gas of open strings with high entropy. Let us consider the moment when the shell is about to pass through the Buchdahl radius. It is easy to calculate the entropy that is available at this point. 
	
	Using $dE=TdS$, and working in a time frame far away from the system where $E=M$ and $T=\frac{8}{81\pi M}$, we recover the standard expression for the entropy given by $S=\pi r^2$, provided that $r=\frac{9M}{4}$.  The tunneling rate is then given by
	\begin{equation}
	\Gamma \sim e^{- \frac{\pi r^2}{4}} e^{\pi r^2} \sim e^{\frac{3\pi r^2}{4}} \,\gg \,1\ .
	\end{equation}
	This suggests that the tunneling is extremely rapid, driven by the huge increase in entropy\footnote{The argument reminds in spirit that of \cite{Kraus:2015zda}, which has been employed in the context of fuzzballs. There, the corresponding tunneling rate was found to be unity.}.  
	To be precise, we should take into account the entropy already present in the matter shell but this will be tiny compared to the one carried by the walls of the final AdS-bubble. 
	
	It is reassuring that in the absence of the entropy available from this collapsing shell of matter the tunneling rate is extremely small, which ensures that the metastable Minkowski vacuum that we live in is extremely long lived and there is no real danger of a spontaneous decay.
	
	Now one might actually wonder whether the tunneling can happen already if the shell has a much larger radius than the Buchdahl radius. This is a more difficult fact to be assessed. 
	The system would be then out of equilibrium, but if it still made sense to associate a temperature with the system, then it should be lower. This could lead to a reduced number of brane/anti-brane pairs, fewer degrees of freedom, and overall a lower entropy.
	At some critical radius, larger than the Buchdahl radius, entropy can no longer compensate for the low tunneling amplitude. If our proposal is correct, one should therefore expect tunneling to occur some time after this critical radius is crossed, but before the Buchdahl radius is reached. If the shell forms at a radius larger than the Buchdahl radius, there will be oscillations and emission of energy before it settles down at the Buchdahl radius.

\section{A stringy realization}\label{sec:string}

	So far we have proposed a 4D effective model capturing some essential features of spherically symmetric black holes and discussed some relevant thermodynamical properties thereof.
	In this last section we further investigate how this may be, first of all, embedded in a 4D supergravity context, and, secondly, we present a concrete stringy setup realizing it.
	
	\subsection{Black shells in SUGRA}
	
		We will now investigate how to realize the above construction within a particular $\mathcal{N}=1$ $D=4$ supergravity coupled to three chiral multiplets inspired from flux compactifications of type II string theory. Consider a spherical bubble with a supersymmetric AdS vacuum in the interior. We label the superpotential inside the bubble by $W_2$. Outside the bubble we consider a no-scale non-supersymmetric Minkowski vacuum, which we label by $W_1$, \emph{i.e.}
		\begin{eqnarray}
		&W_2 \neq 0\ ,&\qquad DW_2 = 0 \ ,\\
		&W_1 \neq 0\ ,&\qquad DW_1 \neq 0\ ,
		\end{eqnarray}
		where $D$ denotes the K\"ahler covariant derivative operator.
		The scalar potential in this $\mathcal{N}=1$ $D=4$ SUGRA is given by
		\begin{equation}
		V \,= \,e^K \left(-3\abs{W}^2+\abs{DW}^2\right)\ ,
		\end{equation}
		where $K$ represents the K\"ahler potential and $W$ is the holomorphic superpotential which we think of as perturbatively induced by fluxes and internal curvature.
		The AdS vacuum inside the bubble is therefore given by $V = -3\abs{W_2}^2 \equiv \Lambda$. This was defined in \fref{sec:shell} in terms of $k$ as $k \equiv \abs{\Lambda}/3$, which gives $\sqrt{k} = \abs{W_2}$.
		
		The shell should have a tension at least as large as the shift in the superpotential across it
		\begin{equation}
		\tau\, \geq \,\frac{\abs{W_2 -W_1}}{4 \pi} \ .
		\end{equation}
		On the other hand, from our solution to the junction conditions in \fref{eq:junction-solution},
		\begin{equation}
		\tau  \,=\, \frac{\abs{W_2}}{4\pi} - \frac{1}{6\pi r}+\frac{1}{16\pi \abs{W_2} r^2}\ ,
		\end{equation}
		where we have used $\sqrt{k} = \abs{W_2}$. For a deep AdS vacuum, the last term is extremely small and can be ignored. The second term is subleading but imposes an upper bound on the tension of the brane
		\begin{equation}
		\frac{\abs{W_2}}{4\pi}\, \geq\, \tau \ .
		\end{equation}
		This gives a bound for the tension of the shell as
		\begin{equation}
		\frac{\abs{W_2}}{4\pi} \,\geq \,\tau \,\geq\, \frac{\abs{W_2-W_1}}{4\pi}\ .
		\end{equation}
		We assume $W_2$ to be the same for all black holes of sufficiently large masses, and all charges. This means that there is a minimum possible value for their size. As we have seen, this will be set by high energy physics and will typically be a few orders of magnitude away from the string scale.
		
		Let us now consider the case of the minimum tension bubble of AdS space inside a pure Minkowski background. Such a shell, however, is not stationary. This can easily be seen by realizing that the junction conditions in \fref{eq:thinshell} do not have a solution when the geometry outside of the bubble is Minkowski (\emph{i.e.} $M=0$). This happens because the junction condition corresponding to energy conservation is solved when the kinetic energy is taken into account, but the junction condition for pressure cannot be satisfied, indicating that there is a net force causing the shell to expand. In a frame of reference that is at rest with respect to the center of the shell, the speed of the expanding shell will approach the speed of light as the radius increases and the shell approaches a flat wall. There is no stationary solution with a flat wall separating AdS from Minkowski spacetime. Luckily, as we have seen in the previous section, the probability of nucleating such an ultra-extremal bubble is very low, and it will take long before the Minkowski space time is destroyed.
	
	\subsection{A model in string theory}
	
		Before moving to the actual stringy realization of the above 4D supergravity model, an important remark is due. In section~\ref{sec:schwarzschild}, we went through the counting of the expected amount of degrees of
		freedom carried by D5-branes wrapped along three compact dimensions and obtained $n\,\sim\,\frac{r}{\ell_{\rm Pl}}$ as a result. 
		Our concrete realization of the shell will actually involve a four-charge brane system dissolved on the shell, where furthermore all four charges will have to be identified in order to correctly reproduce a 4D Reissner-Nordstr{\"o}m black hole. It is therefore natural to require the size of the internal dimensions to be such that the tension of all four branes carrying the charges are the same. In addition, the tensions of the four branes into which they polarize should also be the same. Furthermore, if these branes contribute a fixed fraction to the tension of the shell it follows that  $n\,\sim\,\frac{r}{\ell_{\rm Pl}}$.
		
		After making this remark, let us now construct a concrete realization of such a system in string theory. We will work in type IIA string theory on $\mathbb{T}^{6}/\left(\mathbb{Z}_{2}\times\mathbb{Z}_{2}\right)$ and 
		comment on type IIB at the end. 
		In this case, we retain three complex scalar fields denoted by $(S,T,U)$. The K\"ahler potential reads
		\begin{equation}
		K\,=\,
		-\log\left(-i\,(S-\bar{S})\right)-3\log\left(-i\,(T-\bar{T})\right)-3\log\left(-i\,(U-\bar{U})\right) \ ,
		\end{equation}
		while the superpotential can be written as 
		\begin{equation}
		W \,=\, a_0-3a_1 U+3a_2 U^2-a_3 U^3-b_0 S+3b_1 SU + 3c_0 T+3 c_1 TU \ ,
		\end{equation}
		where the one-to-one relationship between the above various superpotential couplings and type IIA fluxes can be read from \fref{tab:Fluxes}.
		\begin{table}
		\begin{center}
		\begin{tabular}{|c||c|c|c|c|c|c|c|c|c|c|}
		\hline
		Type IIA fluxes & $W$ couplings \\
		\hline\hline
		$F_{(0)}$ & $a_{3}$ \\
		\hline
		$F_{(2)}$ & $a_{2}$ \\
		\hline
		$F_{(4)}$ & $a_{1}$ \\
		\hline
		$F_{(6)}$ & $a_{0}$ \\
		\hline
		$H_{(3)}$ & $b_{0}$ \\
		\hline
		$H_{(3)}$ & $c_{0}$ \\
		\hline
		$\omega$ & $b_{1}$  \\
		\hline
		$\omega$ & $c_{1}$  \\
		\hline
		\end{tabular}\caption{{\it The dictionary between type IIA fluxes and superpotential couplings in compactifications on a twisted $\mathbb{T}^{6}/\left(\mathbb{Z}_{2}\times\mathbb{Z}_{2}\right)$ with R-R \& NS-NS fluxes, 
		    as well as well including metric flux $\omega$. Repeated fluxes may have different independent components inducing different superpotential terms.}}\label{tab:Fluxes}
		\end{center}
		\end{table}
		Let us now consider a no-scale Minkowski background with the superpotential given by
		\begin{equation}
		W_1 \,=\, \sqrt{k}\left(3b_1U^2+b_0 U^3-b_0S+3b_1 SU\right)\ ,
		\end{equation}
		where $k$ is a normalization that will become relevant in the following. In this background, we place a shell composed of the branes in \fref{tab:branes-shell}.
		\begin{table}
		\begin{center}
		\begin{tabular}{|c||c|c|c|c|c|c|c|c|c|c|}
		\hline
		\rule[1em]{0pt}{0pt} -- &$t$ & $\xi _1$ & $\xi _2$ & $r$ &  $x_1$ & $x_2$  & $x_3$  & $x_4$ & $x_5$  & $x_6$  \\
		\hline
		\hline
		\rule[1em]{0pt}{0pt} D8 & $\bigotimes$ & $\bigotimes$  & $\bigotimes$ & -- & $\bigotimes$ & $\bigotimes$ & $\bigotimes$ & $\bigotimes$ & $\bigotimes$ & $\bigotimes$  \\\hline
		\rule[1em]{0pt}{0pt} D4 & $\bigotimes$ & $\bigotimes$  & $\bigotimes$ & -- & $\bigotimes$ & $\bigotimes$ & -- & -- & -- & -- \\\hline
		\rule[1em]{0pt}{0pt} D4 & $\bigotimes$ & $\bigotimes$  & $\bigotimes$ & -- & -- & -- & $\bigotimes$ & $\bigotimes$ & -- & -- \\\hline
		\rule[1em]{0pt}{0pt} D4 & $\bigotimes$ & $\bigotimes$  & $\bigotimes$ & -- & -- & -- & -- & -- & $\bigotimes$ & $\bigotimes$ \\\hline
		\hline
		\rule[1em]{0pt}{0pt} NS5 & $\bigotimes$ & $\bigotimes$  & $\bigotimes$ & -- & $\bigotimes$ & -- & $\bigotimes$ & -- & $\bigotimes$ & --  \\\hline
		\hline
		\rule[1em]{0pt}{0pt} NS5 & $\bigotimes$ & $\bigotimes$  & $\bigotimes$ & -- & -- & -- & -- & $\bigotimes$ & $\bigotimes$ & $\bigotimes$  \\\hline
		\rule[1em]{0pt}{0pt} NS5 & $\bigotimes$ & $\bigotimes$  & $\bigotimes$ & -- & $\bigotimes$ & $\bigotimes$ & -- & -- & -- & $\bigotimes$  \\\hline
		\rule[1em]{0pt}{0pt} NS5 & $\bigotimes$ & $\bigotimes$  & $\bigotimes$ & -- & -- & $\bigotimes$ & $\bigotimes$ & $\bigotimes$ & -- & --  \\\hline
		\hline
		\hline
		\rule[1em]{0pt}{0pt} KK5 & $\bigotimes$ & $\bigotimes$  & $\bigotimes$ & -- & $\bigotimes$ & -- & $\bigotimes$ & -- & iso & $\bigotimes$  \\\hline
		\rule[1em]{0pt}{0pt} KK5 & $\bigotimes$ & $\bigotimes$  & $\bigotimes$ & -- & $\bigotimes$ & $\bigotimes$ & iso & -- & $\bigotimes$ & --  \\\hline
		\rule[1em]{0pt}{0pt} KK5 & $\bigotimes$ & $\bigotimes$  & $\bigotimes$ & -- & iso & -- & $\bigotimes$ & $\bigotimes$ & $\bigotimes$ & --  \\\hline
		\hline
		\rule[1em]{0pt}{0pt} KK5 & $\bigotimes$ & $\bigotimes$  & $\bigotimes$ & -- & -- & $\bigotimes$ & -- & $\bigotimes$ & iso & $\bigotimes$ \\\hline
		\rule[1em]{0pt}{0pt} KK5 & $\bigotimes$ & $\bigotimes$  & $\bigotimes$ & -- & -- & $\bigotimes$ & iso & $\bigotimes$ & -- & $\bigotimes$ \\\hline
		\rule[1em]{0pt}{0pt} KK5 & $\bigotimes$ & $\bigotimes$  & $\bigotimes$ & -- & iso & $\bigotimes$ & -- & $\bigotimes$ & -- & $\bigotimes$ \\\hline
		\hline
		\rule[1em]{0pt}{0pt} KK5 & $\bigotimes$ & $\bigotimes$  & $\bigotimes$ & -- & -- & $\bigotimes$ & $\bigotimes$ & -- & $\bigotimes$ & iso \\\hline
		\rule[1em]{0pt}{0pt} KK5 & $\bigotimes$ & $\bigotimes$  & $\bigotimes$ & -- & $\bigotimes$ & -- & -- & iso & $\bigotimes$ & $\bigotimes$ \\\hline
		\rule[1em]{0pt}{0pt} KK5 & $\bigotimes$ & $\bigotimes$  & $\bigotimes$ & -- & $\bigotimes$ & iso & $\bigotimes$ & -- & -- & $\bigotimes$ \\\hline
		\end{tabular}
		\caption{{\it Arrangement of branes comprising the shell. Each brane in this system realizes a jump of the corresponding flux when going across the shell.}}
		\label{tab:branes-shell}
		\end{center}
		\end{table}
		We want to construct a supersymmetric AdS vacuum inside the shell for which we pick the solution from \cite{Dibitetto:2011gm} given by
		\begin{eqnarray}
		W_2 &=& \sqrt{k} \bigg(\frac{3\sqrt{10}}{2} -\frac{3\sqrt{6}}{2}U -\frac{\sqrt{10}}{2}U^2 -\frac{5}{\sqrt{6}} U^3 +\frac{\sqrt{6}}{3} S \nonumber\\
		&+& \sqrt{10} SU + \sqrt{6} T +3 \sqrt{10} TU\bigg)\ ,
		\end{eqnarray}
		where $k$ is given in terms of the AdS vacuum as before. The difference in the superpotential across the shell should be generated by shifts in the fluxes associated with a brane whose tension must obey $\tau \geq |\Delta W|/4\pi$. Our goal therefore is to choose parameters such that this brane be composed of the branes listed it \fref{tab:branes-shell}.
		
\begin{figure}
		\begin{center}
		\begin{tabular}{cc}
		\includegraphics[width=0.5\textwidth]{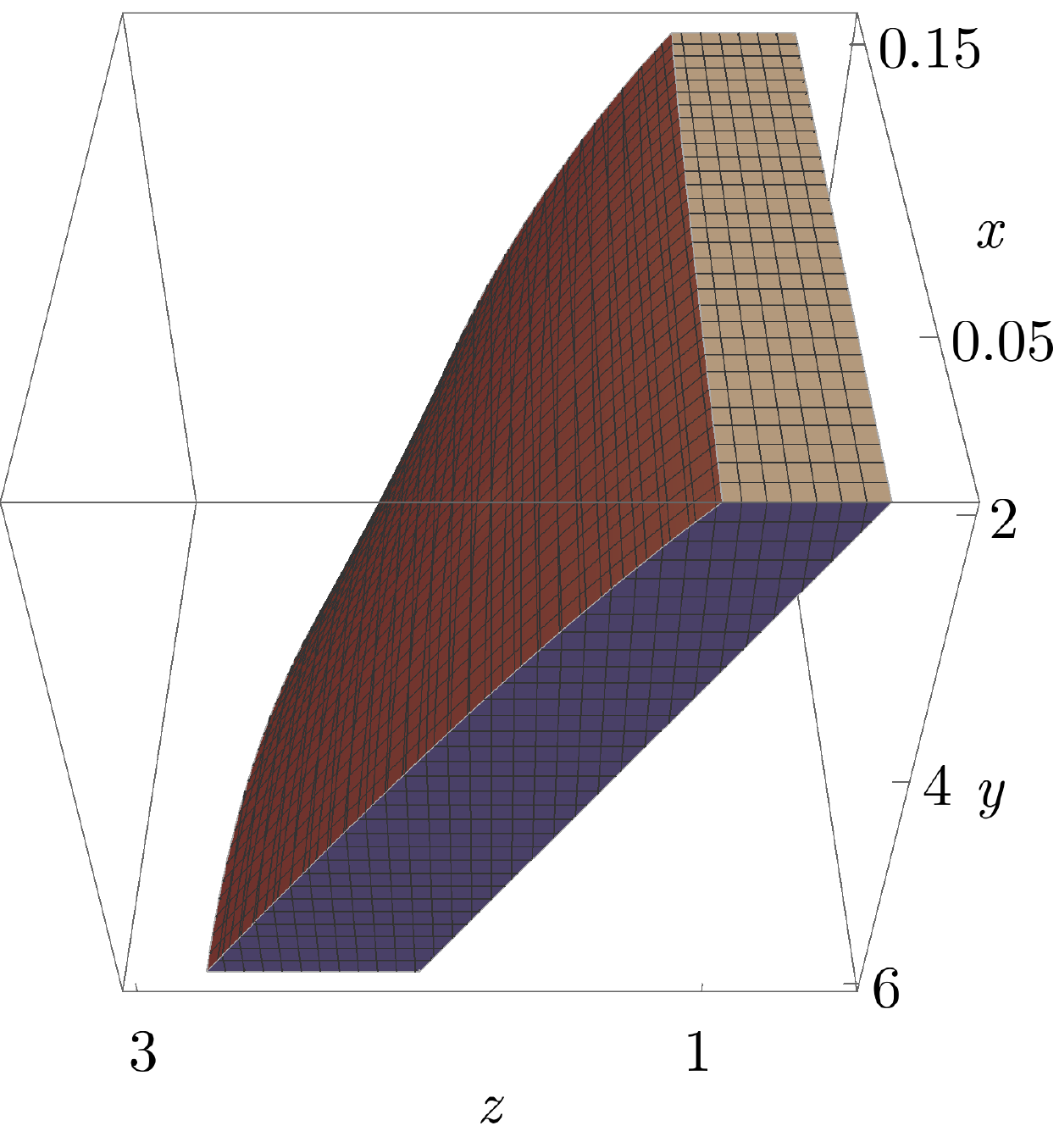}  &  \includegraphics[width=0.5\textwidth]{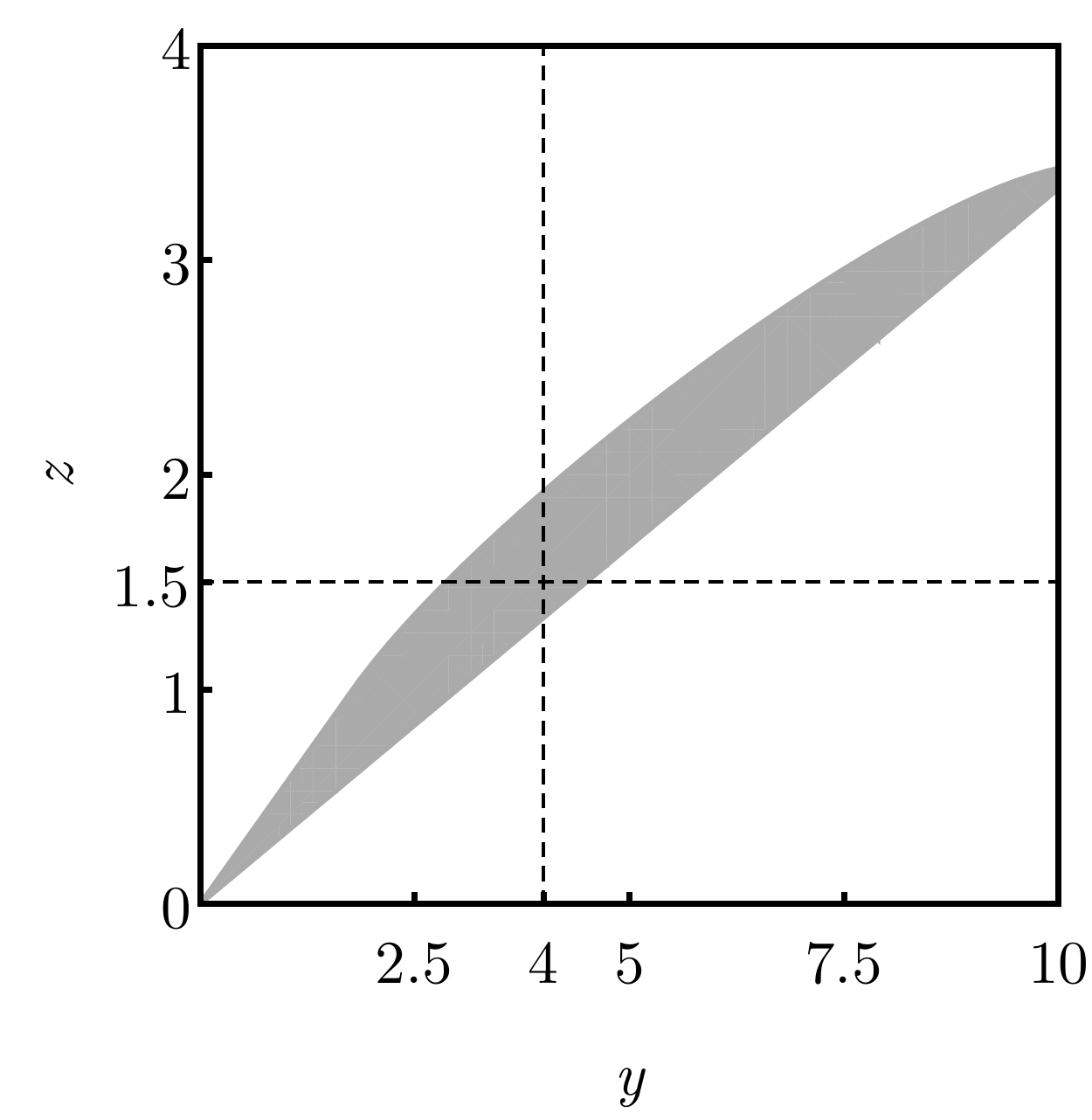}
		\end{tabular}
		\caption{{\it \emph{Left:} The region of the three-dimensional parameter space corresponding to valid explicit realizations of flux shifts compatible with a bubble wall of the type sketched in table~\fref{tab:branes-shell}. 
		    \emph{Right:} A two-dimensional slice of the parameter space on the left corresponding to the explicit choice $x=\frac{1}{10}$.}}
		\label{fig:parameter}
		\end{center}
		\end{figure}

		To achieve this, we need to scale the moduli and move away from the origin of the moduli space by turning on axions and read off the corresponding fluxes. We do the following non-compact $\textrm{SL}(2)^{3}$
		transformations
		\begin{eqnarray}
		S &\mapsto& x~S + \tilde{x}\ , \\
		T &\mapsto& y~T + \tilde{y}\ , \\
		U &\mapsto& z~U + \tilde{z}\ ,
		\end{eqnarray}
		where the shifts $\left( \tilde{x},\tilde{y},\tilde{z} \right)$ are given in terms of the rescaling parameters $\left(x,y,z \right)$ by
		\begin{equation}
		\begin{split}
		\tilde{x} \,=& \,\frac{3y+2z-D}{8\sqrt{15}z}\ , \\
		\tilde{y} \,=& \,\frac{-D \left(\left(37-36 x^2\right) y z+2 \left(54 x^2+11\right) z^2+3 y^2\right)+3 \left(36 x^2-49\right) y^2 z}{96 \sqrt{15} z^2 (y-3 z)}\\
		&\,\frac{-4 \left(63x^2+187\right) y z^2+4 \left(71-54 x^2\right) z^3+9 y^3}{96 \sqrt{15} z^2 (y-3 z)}\ ,\\
		\tilde{z} \,=& \,\frac{D \left(\left(67-12 x^2\right) y z+2 \left(18 x^2-7\right) z^2-3 y^2\right)+3 \left(71-12 x^2\right) y^2 z}{96 \sqrt{15} y z (y-3 z)}\\
		&\,\frac{+28 \left(3
		  x^2+13\right) y z^2+4 \left(18 x^2-53\right) z^3+9 y^3}{96 \sqrt{15} y z (y-3 z)}\ ,
		\end{split}
		\end{equation}
		with $D$ defined as
		\begin{equation}\label{eq:D}
		D \,\equiv\, \sqrt{9y^2 + 252 y z - 476 z^2}\ .
		\end{equation}
		These shifts in the moduli, shift the superpotentials $W_1$ and $W_2$, and their difference $\Delta W \equiv W_2-W_1$, in such a way that it is possible to satisfy the requirements of symmetry outlined in the beginning of the section. In particular we want a shift that is symmetric in all the fluxes of the form
		\begin{equation}
		\abs{\Delta a_1} \,=\, \abs{\Delta a_3},\qquad \abs{\Delta b_0} \,= \,\abs{\Delta c_0},\qquad \abs{\Delta b_1} \,= \,\abs{\Delta c_1}\ .
		\end{equation}
		Furthermore, our construction does not contain branes sourcing $\Delta a_0$ or $\Delta a_2$ on the shell and so we need them to vanish in $\Delta W$.
		This determines $b_0$ and $b_1$ in terms of $\left(x,y,z \right)$ as
		\begin{equation}
		\begin{split}
		b_0 \,&=\, \frac{2z - 7y + D}{2 \sqrt{6}y}\ ,\\
		b_1 \,&= \,\frac{\sqrt{10}\left(y-3z\right)}{3y}\ ,
		\end{split}
		\end{equation}
		where $D$ is the quantity defined in \fref{eq:D}.
		In order to have ultracriticality \emph{i.e.}
		\begin{equation} \label{eq:w_delw_ineq}
		\frac{\abs{W_2}}{4 \pi} \,\geq \,\frac{\abs{W_2 -W_1}}{4 \pi}\ ,
		\end{equation}
		the parameters $\left(x,y,z\right)$ must lie in the region shown in \fref{fig:parameter}.
		Picking a point in this region, for concreteness  $x=1/10,y=4,z=3/2$, 
		the superpotential outside and inside the bubble after shifts and rescaling in moduli space read
		\begin{equation}
		\begin{split}
		W_2 \,= \,\sqrt{k}&\left(2 \sqrt{\frac{2}{5}} S U-\sqrt{\frac{2}{3}} \left(\sqrt{65}-9\right) S+\frac{9}{2 \sqrt{10}}TU-\frac{3}{4} \sqrt{\frac{3}{2}} \left(\sqrt{65}-9\right) T\right.\\
		&-\frac{U^3}{200 \sqrt{6}}-\frac{U^2}{10 \sqrt{5 \left(73+9
		    \sqrt{65}\right)}}-\frac{\left(26245+1221 \sqrt{65}\right)}{9600 \sqrt{6}}U\\
		&\left.+\frac{3683 \sqrt{10}+367 \sqrt{26}}{2880}\right),
		\end{split}
		\end{equation}
		and
		\begin{equation}
		\begin{split}
		W_1\, =\, \sqrt{k}&\left(-\frac{S U}{2 \sqrt{10}}-\frac{5 \left(\sqrt{65}-9\right)}{4 \sqrt{6}}S+\frac{\left(3 \sqrt{65}-25\right)}{8000 \sqrt{6}}U^3\right.\\
		&-\frac{U^2}{10 \sqrt{5 \left(73+9 \sqrt{65}\right)}}-\frac{\left(131495+6159 \sqrt{65}\right)}{48000 \sqrt{6}}U\\
		&\left.+\frac{3683 \sqrt{10}+367 \sqrt{26}}{2880}\right),
		\end{split}
		\end{equation}
		respectively.
		This gives the jump in the superpotential across the shell as
		\begin{equation}\label{eq:deltaW}
		\Delta W \, =\, 3aU -aU^3 +bS +3bT +cSU +cTU\ ,
		\end{equation}
		where
		\begin{equation}
		\begin{split}
		a \,&= \,\sqrt{\frac{3}{2}}\frac{\left(5+\sqrt{65}\right)}{8000}\sqrt{k}\ ,\\
		b \,&= \,\frac{1}{4} \sqrt{\frac{3}{2}} \left(9-\sqrt{65}\right)\sqrt{k}\ ,\\
		c \,&= \,\frac{9}{2 \sqrt{10}}\sqrt{k}\ .
		\end{split}
		\end{equation}
		Evaluated numerically this is
		\begin{equation}
		\Delta W \,= \,\sqrt{k}\left(0.006 U - 0.002 U^3 + 0.29 S + 0.87 T + 1.42 SU + 1.42 TU\right)\ .
		\end{equation}
		Evaluated at the origin of moduli space ($S=T=U=i$), the real and imaginary parts of the jump in the superpotential add with the same signs respectively as expected. The inequality \fref{eq:w_delw_ineq} can also be checked explicitly
		\begin{equation} 
		6.5 \sqrt{k} \,> \,3.1 \sqrt{k}\ . 
		\end{equation}
		
		Now that we have a concrete construction of the shell in string theory, let us take a moment to understand the underlying geometry. The first two terms in \fref{eq:deltaW} come from the D8-D4-D4-D4 system. The next two terms reflect the presence of a singlet NS5 and a triplet of NS5-branes on the shell. The fluxes corresponding to all of these branes contribute to the imaginary part of the superpotential and come with the same sign. There is no binding energy between these branes. 
		
		The last two terms come from the KK5-monopoles and contribute to the real part of the jump in the superpotential. They are responsible for a change in the metric flux. They also add with the same relative sign, and have no binding energy between them.
		
		There is, however, binding energy between the KK5-monopoles and the other branes. This can be seen when computing $\,\abs{\Delta W} = \sqrt{\left(\mathrm{Re}\left(\Delta W\right)\right)^2 + \left(\mathrm{Im}\left(\Delta W\right)\right)^2}$, and can be understood as  the KK5-monopoles binding the other branes together and preventing them from drifting away.
		
		To summarize, branes that make up the shell are
		\begin{enumerate}[(i)]
		\item a D8-brane sourcing an $F_{(0)}$ flux,
		\item a triplet of D4-branes sourcing three different $F_{(4)}$ fluxes, 
		\item NS5-branes sourcing 3-form fluxes $H_{(3)}$,
		\item KK-monopoles sourcing metric fluxes $\omega$.
		\end{enumerate} 
		
		After constructing the shell, let us now construct a Reissner-Nordström black hole from branes dissolved in the shell. We consider a system composed by the branes in \fref{tab:branes-RN}.
		\begin{table}
		\begin{center}
		\begin{tabular}{|c|c|c|c|c|c|c|c|c|c|c|}
		\hline
		\rule[1em]{0pt}{0pt} -- &t & $\xi _1$ & $\xi _2$ & r &  $x^1$ & $x^2$  & $x^3$  & $x^4$ & $x^5$  & $x^6$  \\
		\hline
		\hline
		\rule[1em]{0pt}{0pt} D6 & $\bigotimes$ & --  & -- & -- & $\bigotimes$ & $\bigotimes$ & $\bigotimes$ & $\bigotimes$ & $\bigotimes$ & $\bigotimes$  \\\hline
		\rule[1em]{0pt}{0pt} D2 & $\bigotimes$ & --  & -- & -- & $\bigotimes$ & $\bigotimes$ & -- & -- & -- & -- \\\hline
		\rule[1em]{0pt}{0pt} D2 & $\bigotimes$ & --  & -- & -- & -- & -- & $\bigotimes$ & $\bigotimes$ & -- & -- \\\hline
		\rule[1em]{0pt}{0pt} D2 & $\bigotimes$ & --  & -- & -- & -- & -- & -- & -- & $\bigotimes$ & $\bigotimes$ \\\hline        
		\end{tabular}
		\caption{\it Dissolved branes producing a Reissner-Nordström black hole}
		\label{tab:branes-RN}
		\end{center}
		\end{table}
		The corresponding 10D metric is given by
		\begin{equation}
		\begin{aligned}
		\mathrm{ds}_{10}^2 \,=\, &-\left( H^{\db6} \prod_{i=1}^{3} H^{\db2}_i \right)^{-1/2} \D t^2 + \left( H^{\db6} \prod_{i=1}^{3} H^{\db2}_i \right) ^{1/2} r^2 \D \Omega_{(2)}^2\\
		&+ \sqrt{\frac{H^{\db2}_{2}H^{\db2}_{3}}{H^{\db6} H^{\db2}_{1}}} \left( (\D x_1)^2+(\D x_2)^2\right)
		+ \sqrt{\frac{H^{\db2}_{1}H^{\db2}_{3}}{H^{\db6} H^{\db2}_{2}}} \left( (\D x_3)^2+(\D x_4)^2\right) \\
		&+ \sqrt{\frac{H^{\db2}_{1}H^{\db2}_{2}}{H^{\db6} H^{\db2}_{3}}} \left( (\D x_5)^2+(\D x_6)^2\right)\ ,
		\end{aligned}
		\end{equation}
		and the dilaton
		\begin{equation}
		e^{2\phi}\, = \,\left( \prod_{i=3}^{3} H^{\db2}_i \right) ^{1/2}\left(  H^{\db6}\right) ^{-3/2}\ .
		\end{equation}
		The R-R potentials read
		\begin{eqnarray}
		C_{(3)} &=& \left[\left(H^{\db2}_{1}\right)^{-1} -1\right] \D t \wedge \D x_1 \wedge \D x_2 \nonumber\\
		&+& \left[\left(H^{\db2}_{2}\right)^{-1} -1\right] \D t \wedge \D x_3 \wedge \D x_4 \\
		&+& \left[\left(H^{\db2}_{3}\right)^{-1} -1\right] \D t \wedge \D x_5 \wedge \D x_6\ , \nonumber\\
		C_{(7)} &=& \left[\left(H^{\db6}\right)^{-1} -1\right] \D t \wedge \D x_1 \wedge \D x_2 \wedge \D x_3 \wedge \D x_4 \wedge \D x_5 \wedge \D x_6\ ,
		\end{eqnarray}
		yielding the following non-vanishing components for the R-R field strengths 
		\begin{equation}
		\begin{array}{lclclclc}
		F_{t r x_1 x_2} & , & F_{t r x_3 x_4} & , & F_{t r x_5 x_6} & ; & F_{t r x_1  x_2  x_3  x_4  x_5  x_6} & .
		\end{array}\nonumber
		\end{equation}
		
		We now put
		\begin{equation}
		H^{\db2}_{1}\,=\,H^{\db2}_{2}\,=\,H^{\db2}_{2}\,=\,H^{\db6}\,= \,\left(1-\frac{Q}{r}\right)^{-1}\ ,
		\end{equation}
		which makes all field strengths equal contributing to $F^{t r} = Q/r$.
		Compactifying directions $x^1,\ldots,x^6$ on a torus $\mathbb{T}^6$, the metric becomes that of a Reissner-Nordström black hole with a constant dilaton. 
		
		The picture of a black hole we have constructed is that of a shell made up of bound system of D8, D4, NS5-branes and KK-monopoles. The four-charge Reissner-Nordström black hole arises as D6-D2-D2-D2 branes that are dissolved in the shell. The size of the extra dimensions, yielding symmetric tensions, obey $L_2=\ell_s^2/L_1$,  $L_4=\ell_s^2/L_3$ , and  $L_6=\ell_s^2/L_5$ . Typically three of these will be smaller than string scale, which can be fixed by three T-dualities to type IIB,  with charges carried by D3-branes. The price to pay is that non-geometric fluxes are needed to support the AdS-vacuum. Furthermore, we expect it to be possible to build non-extremal solutions on this taking inspiration from \cite{Danielsson:2001xe}.

\section{Conclusions}\label{sec:conclusions}

	In this paper we have investigated an alternative to black holes in the form of gravastars built out of branes surrounding bubbles of AdS-space. We have argued that such configurations not only solve the equations of motion, but are also stable against small perturbations when the thermodynamical properties are taken into account. Crucial to our construction, is that the background Minkowski vacuum is non-perturbatively unstable towards a transition into the AdS-vacuum. The decay time is shown to be large, and the Minkowski vacuum sufficiently long lived. We also show that a collapsing shell of matter will initiate a rapid transition when it becomes smaller than some critical radius. This leads to a bubble stabilized at a final radius larger than the horizon radius of the black hole that otherwise would have formed. In case of a neutral black hole this turns out to be the Buchdahl radius at $r=\frac{9M}{4}$.
	
	If the black hole is charged, the radius will be smaller. In the limit of extremal Reissner-Nordst{\"o}m, the shell approaches the horizon. Even though the asymptotic temperature is zero, as expected, the local temperature is finite at the position of the shell due to the diverging blueshift. Hence, the otherwise somewhat mysterious non-zero entropy of a zero temperature extremal black hole can ultimately be traced to a gas at finite temperature.
	
	In the paper we manage to identify all the necessary building blocks within string theory. This involves a set of branes building up the shell, together with lower dimensional dissolved branes, as well as a gas of open strings. We go through a detailed example in type IIA, where we explicitly identify all the different kind of branes, and show how the fluxes shift between the vacuum outside and inside of the shell. It is highly non-trivial that we are able to satisfy the necessary requirements to form stable gravastars. 
	
	Our analysis suggests that there are ultra compact objects in string theory that from far away look very similar to black holes, and that these, rather than proper black holes, are the likely result of gravitational collapse. The proposal has many similarities in spirit with the fuzzballs, but is in many respects different. In particular, the anatomy of the object is different, with a thin shell enclosing empty AdS space rather than a fuzzy clump of matter. 
	
	When generalized to rotating systems, our results should be relevant to the recent observations of gravitational radiation from colliding black holes. For instance, it has been proposed that hard surfaces outside of the horizon could give rise to echoes of gravitational waves, see  \cite{Abedi:2016hgu,Cardoso:2016oxy,Cardoso:2016rao,Barcelo:2017lnx}. Similarly, while two colliding {\it bona fide} black holes only emit gravitational radiation, it is less clear what happens in the case of two shells. Furthermore, it would be interesting to consider what kind of signatures one should look for with the Event Horizon Telescope. Still, it is likely that any effect of this sort will be heavily suppressed. Our shells have an enormous number of degrees of freedom, and entropy considerations suggest that infalling matter is absorbed with an extreme efficiency. This is also the way that the shells may evade the kind of bounds discussed in \cite{Broderick:2009ph}. On the other hand, given the size of our shells, there could be non-trivial modifications of the ring down signal. We hope to return to these questions in the near future.

\section*{Acknowledgments}

	We would like to thank Vishnu Jejjala, Dietmar Klemm, Nicolò Petri, Thomas Van Riet, Sergio Vargas and Bert Vercnocke for stimulating discussions.
	The work of the authors was supported by the Swedish Research Council (VR). 

\small
\bibliography{references}
\bibliographystyle{utphys}

\end{document}